\documentclass[prl,twocolumn,superscriptaddress,preprintnumbers,amsmath,amssymb]{revtex4}
\usepackage[dvips]{graphicx}
\usepackage{dcolumn}
\usepackage{color}
\usepackage{bm}

\usepackage{epstopdf}

\newcommand{\work}{ \langle W \rangle }
\newcommand{\Wsquare}{\langle W^2 \rangle}
\newcommand{\Wvariance}{\langle \delta W^2 \rangle}
\newcommand{\power}{\hat{P}}

\newcommand{\hatU}{\hat{U}}
\newcommand{\Trace}{{\rm Tr}}

\newcommand{\twobytwo}[4]{\left( \begin{array}{cc} #1 & #2 \\ #3 & #4 \end{array}\right)}
\newcommand{\phiS}{\varphi}
\newcommand{\phid}{\xi}

\newcommand{\hatL}{\hat{\mathcal{L}}}

\newcommand{\hatH}{\hat{H}}
\newcommand{\hatn}{\hat{n}}

\newcommand{\hatrho}{\hat{\rho}}

\newcommand{\LZ}{P_{LZ}}
\newcommand{\Time}{\mathcal{T}}

\def\braket#1{\mathinner{\langle{#1}\rangle}}

\begin{document}

\title{Work and its fluctuations in a driven quantum system}

\author{Paolo Solinas}
\affiliation{Low Temperature Laboratory (OVLL), Aalto University School of Science, P.O. Box 13500, 00076 Aalto, Finland}
\affiliation{Department of Applied Physics, Aalto University School of Science,
P.O. Box 11000, 00076 Aalto, Finland}
\author{Dmitri V.\ Averin}
\affiliation{Department of Physics and Astronomy, Stony Brook University, SUNY, Stony Brook, NY 11794-3800,
USA}
\author{Jukka P. Pekola}
\affiliation{Low Temperature Laboratory (OVLL), Aalto University School of Science, P.O. Box 13500, 00076 Aalto, Finland}

\begin{abstract}
We analyze work done on a quantum system driven by a control field. The average work depends on the whole dynamics of the system, and is obtained as the integral of the average power operator. As a specific example we focus on a superconducting Cooper-pair box forming a two-level system. We obtain expressions for the average work and work distribution
in a closed system, and discuss control field and environment contributions to the average work for an open system. 
\end{abstract}

\date{\today}

\maketitle
The fluctuation relations (FRs) \cite{bochkov77,jarzynski97} govern work and dissipation in small classical systems when they are driven out of equilibrium. They have recently attracted lots of attention because of their applications in molecular systems~\cite{alemany11}.
Fluctuation relations can also be accurately studied in single-electron transport \cite{averin11, kung12, saira12}.
A natural question is if similar concepts and experiments can be extended to the quantum regime.
The first attempts in this direction focused on finding a proper work operator \cite{bochkov77, yukawa00, chernyak04, allahverdyan05, engel07}.
However, after a long debate, it has become clear that this approach has serious drawbacks \cite{talkner07}.
Work is characterized by a process, not only by the instantaneous state of the system \cite{talkner07,campisi11}, and therefore it cannot be defined by an operator local in time, which would disregard the actual evolution of the system under the driving protocol.
Although this is not an issue for closed systems it can become critical when discussing work in open systems.
Alternatively the work has been defined through a two-measurement process (TMP) \cite{kurchan00, tasaki00, mukamel03, talkner07, engel07, esposito09, campisi11}. The energy of the system is measured at the beginning and at the end of the evolution and the work done in a process is determined by the corresponding energy difference.
This definition has the advantage that the quantum FRs can be immediately obtained and they resemble the classical ones.
In this proposal the system does not interact with the environment and, thus, the dynamics is unitary.

To circumvent the problem of extending the TMP approach to an open system~\cite{campisi09,talkner09}, we introduce work in analogy to that in the classical case as an integral of the injected power during the evolution.
Let the evolution of the system be governed by a time-dependent Hamiltonian $\hatH(t)$ driven by a control parameter $\lambda(t)$.
The corresponding power operator is then given by $\power =  \partial \hatH/\partial \lambda  \dot{\lambda} =\partial \hatH/\partial t$.
If the state of the system is described by its reduced density operator $\hat \rho(t)$, the average power is given by $\braket{\power(t)} = \Trace \{ \hatrho(t) \power(t) \}$ and the expectation value of the work done on the quantum system is
\begin{equation}
\work = \int_0^\Time \langle \power(t) \rangle dt.
 \label{eq:work}
\end{equation}
This way, the work explicitly depends on the whole evolution of the system through $\hatrho(t)$ containing the information about the dynamics which can be unitary or not.
To address this point, we
differentiate the average energy of the {\sl system}, $\braket{\hatH} = \Trace \{ \hatrho(t) \hatH(t) \}$, yielding 
\begin{equation}
\frac{d}{dt}\braket{\hatH}  
= \Trace \{ \frac{d\hatrho}{dt} \hatH \} + \braket{\power}
\label{eq:dH}
\end{equation}
Under quite general assumptions the dynamics of the reduced density operator of the system can be described by a master equation \cite{breuer} $d \hatrho/dt= -\frac{i}{\hbar} [\hatH,\hatrho] + \hatL(\rho)$
where the contributions on the right-hand-side are given by the unitary and dissipative dynamics, respectively.
By substituting the above result into Eq. (\ref{eq:dH}), we find that there is no contribution due to the unitary dynamics since $\Trace \{ [\hatH,\hatrho] \hatH \}$ vanishes.
Then the average power reads
$\braket{\power(t)}=  d\braket{\hatH(t)}/dt - \Trace \{ \hatL(\rho)  \hatH(t) \}$ and the corresponding average work is given by
\begin{equation}
\work =   \langle \hatH(\Time) \rangle - \langle \hatH(0) \rangle - \int_0^\Time dt\Trace  \{  \hatL(\rho)  \hatH(t)  \}.
\label{eq:work_dynamics}
\end{equation}
If the system does not interact with the environment, only the first difference on the right-hand-side survives in Eq. \eqref{eq:work_dynamics}, and the average work is equal to the variation of the internal energy.
The last term describes the energy exchange with the environment during the evolution process, and is dependant on the particular realization of the evolution trajectory. We call this contribution average heat and denote it as $Q$.
In thermodynamical terms, Eq. (\ref{eq:work_dynamics}) is the first law in the quantum regime, and it has been discussed previously in Refs. \cite{allahverdyan01,esposito06} as the energy balance equation. 

The average work definition in Eq. (\ref{eq:work_dynamics}) is more general than the TMP since it takes into account the full quantum evolution. For a closed quantum system, given the initial quantum state and a driving protocol, the evolution is determined completely by the Schr\"odinger equation, while the initial measurement performed in the TMP causes a collapse of the quantum state before the beginning of the protocol.
Thus, the two approaches yield different results for the initial states with coherent superpositions of the eigenstates of $ \hatH(0)$ and the TMP result is recovered if the system is initially in an eigenstate or in an incoherent superposition of the eigenstates of $ \hatH(0)$.

The second advantage of Eq. (\ref{eq:work_dynamics}) is that it allows to define and calculate the heat only in terms of the system density matrix (and its dynamics) and, thus, it avoids the formidable task of measuring the environment degrees of freedom as in the TMP \cite{kurchan00, campisi11}.
In addition, no assumption has been made on the dissipative dynamics then Eq. (\ref{eq:work_dynamics}) can be used with any master equation (Lindblad, non-Lindblad, non-Markovian).



{\it Cooper-pair box as a driven quantum two-level system}. We consider a Cooper-pair box (CPB) \cite{averin85,bouchiat98, nakamura99} consisting of a superconducting island connected to a superconducting lead by a Josephson tunnel junction.
The system is described by the circuit scheme in the inset in Fig. \ref{fig:Work_and_Variance} and it is characterized by a voltage source $V_g$, coupling gate capacitance $C_g$, a Josephson junction with energy $E_J$ and capacitance $C_J$. We denote $C_\Sigma \equiv C_g + C_J$. Resistor $R$, to be discussed in the last part of the paper, forms the dissipative environment of the box.
This system is a good candidate for a calorimetric measurement of quantum work distribution \cite{pekola12}.

In the regime $\epsilon \equiv E_J/(2E_C) \ll 1$, where $E_C=2 e^2/C_\Sigma$ is the charging energy of the box, we can treat the CPB as a two-level quantum system.
Denoting with $|0\rangle $ and $|1\rangle $ the state with zero and one excess Cooper-pairs on the island, respectively, the Hamiltonian reads
\begin{equation}
 \hatH=  -E_C q (|1\rangle \langle 1| -|0\rangle \langle 0|)- \frac{E_J}{2} (|1\rangle \langle 0| +|0\rangle \langle 1|),
 \label{eq:CPB_Ham}
\end{equation}
where $q= C_g V_g/(2e)-1/2$ is the normalized gate voltage.
We assume driven evolution: a linear gate ramp $q(t)=-1/2+ t/\Time$ over a period $\Time$ starting from  $t=0$.
The ground and the excited states of the system are separated by the energy gap $\hbar \omega_0 = 2 E_C \sqrt{q^2+\epsilon^2}$ 
which reaches the minimum $E_J$ at $t=\Time/2$, see Fig. \ref{fig:spectrum}.
We recover the standard Landau-Zener (LZ) model \cite{zener32, campisi_PRE11} where the system is excited when driven in a non-adiabatic way through a avoided crossing.
The time-dependent eigenstates of the Hamiltonian (\ref{eq:CPB_Ham}) are $|g\rangle = \frac{1}{\sqrt{2}}( \sqrt{1-\eta} |0\rangle +  \sqrt{1+\eta} |1\rangle  )$ and $|e\rangle = \frac{1}{\sqrt{2}}(\sqrt{1+\eta} |0\rangle - \sqrt{1-\eta} |1\rangle  )$
where $\eta = q/\sqrt{q^2+\epsilon^2}$ \cite{solinasPRB10}.

For this system, the power operator is $\power =   E_C \dot q( \openone - 2  \hatn)$, where $\hatn = |1\rangle \langle 1|$ is the operator of the number of Cooper pairs on the island and $\openone = |1\rangle \langle 1| + |0\rangle \langle 0|$ is the identity operator.
We calculate the time-dependent average of the power operator in the Heisenberg picture, $\power^H(t) = U^\dagger(t) \power U(t)$, with the time evolution operator $U(t)$, and the state $|\psi (0)\rangle$ that does not change in time. 
Here we focus on the first and second moments of the work done on the CPB, which can be expressed through $\power^H(t)$ as 
$\work =  \int_0^\Time dt \langle \power^H(t) \rangle$ and
$\Wsquare=  \int_0^\Time dt_2 \int_0^{\Time} dt_1\langle \power^H(t_2) \power^H(t_1) \rangle$, where $\langle ... \rangle \equiv  \langle \psi (0)| ... |\psi (0)\rangle$.
Explicitly: 
\begin{equation}
 	\work =  E_C \Big(1 - \frac{2}{\Time} \int_0^\Time \langle \hatn^H(t) \rangle dt \Big)
	\label{eq:work_CPB}
\end{equation}
and
\begin{eqnarray}  \label{eq:work_variance}
 &&\Wsquare  =  2 E_C \work \\ &&
 -  E_C^2 \Big[1 - \frac{4}{\Time^2} \int_0^\Time dt_2 \int_0^{\Time} dt_1   \langle \hatn^H(t_2) \hatn^H(t_1)\rangle   \Big]\nonumber.
\end{eqnarray}
Equations (\ref{eq:work_CPB}) and (\ref{eq:work_variance}) can be applied for both closed and open systems~\cite{breuer,gardiner}.
\begin{figure}
    \begin{center}
    \includegraphics[scale=.35]{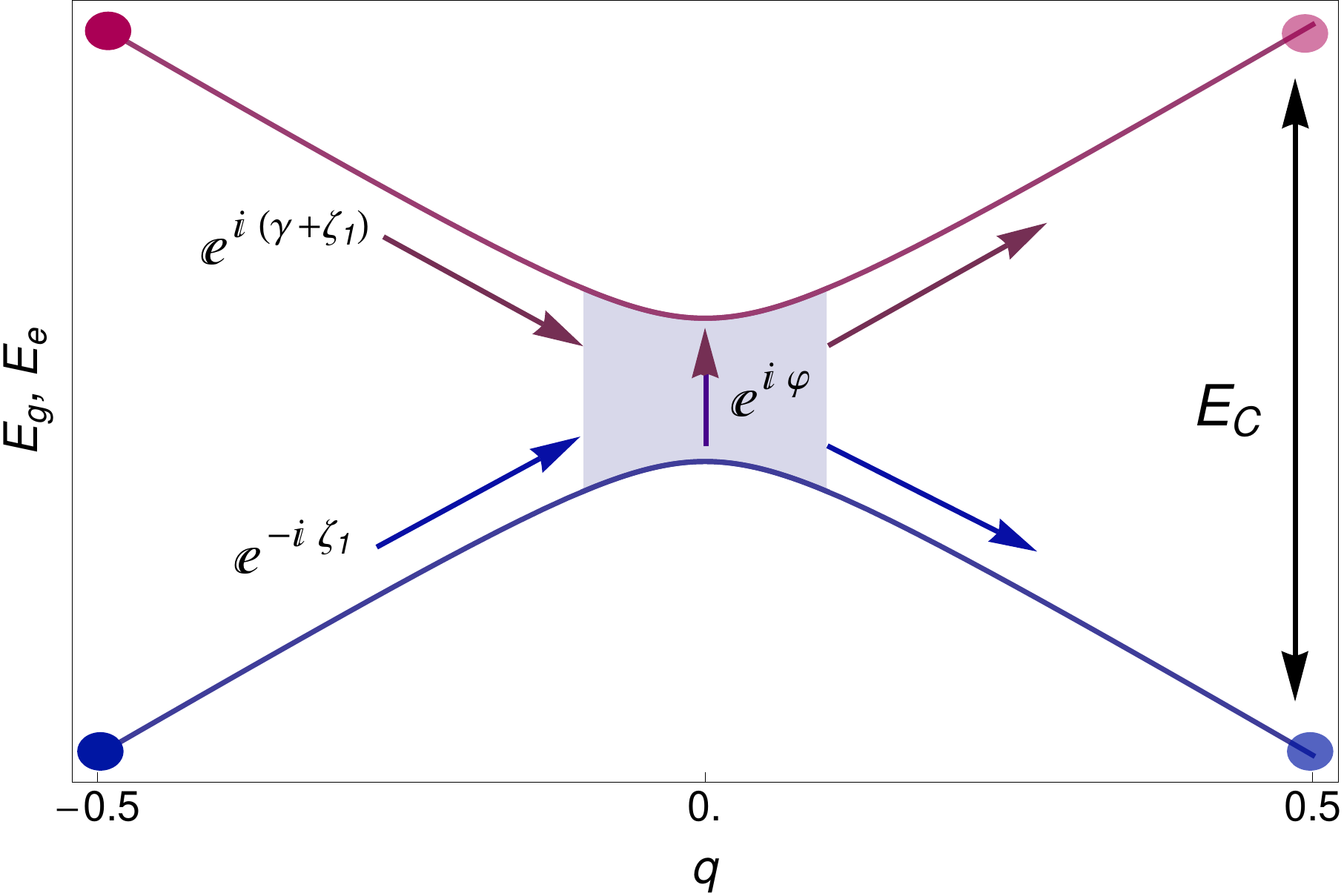}
   \end{center}
    \caption{(Color online) Schematic presentation of avoided crossing with the eigenstates of energies $E_g$ and $E_e$ in a CPB as a function of the normalized gate charge $q$. 
The phases which contribute to the interference are explicitly indicated.}
    \label{fig:spectrum}
\end{figure}

{\it Instantaneous transition regime, unitary evolution}.
If the time of the control ramp is much shorter than the relaxation and dephasing times, the evolution of the system can be considered unitary.
For $\epsilon \ll 1$, the LZ transitions are localized near the minimum energy gap at $t=\Time/2$ and the dynamics is well approximated by the instantaneous transition model \cite{shevchenko10, gasparinetti11}, i.e., the evolution is composed of pure adiabatic evolution and instantaneous LZ transitions at $t=\Time/2$, see Fig. \ref{fig:spectrum}. All work, spent exactly in these LZ transitions, is stored in the system (CPB) as increased internal energy.
Along the adiabatic region, the evolution operator reads $U_k(t)= \exp{[- i \phid_k^t \sigma_z]}$ where $\phid_k^t=\int_{t_{k-1}}^{t_{k}} d\tau \omega_0 (\tau)/2$ is half of the integrated energy gap, $t_k$ and $t_{k-1}$ denote the time limits of the adiabatic evolution and $\sigma_z =  |e(t)\rangle \langle e(t)| - |g(t)\rangle \langle g(t)|$.
The transfer matrix  for the instantaneous LZ transitions in the basis $\{|g(\Time/2)\rangle , |e(\Time/2)\rangle  \}$ reads
\begin{equation}
N_{LZ} =\twobytwo{\sqrt{1-\LZ} e^{i\phiS}}{- \sqrt{\LZ}}{\sqrt{\LZ}}{\sqrt{1-\LZ} e^{-i\phiS}}\ ,
\end{equation}
where $\phiS= \delta (\log\delta -1 ) + \arg \Gamma (1-i\delta) - \pi/4$ 
($\Gamma$ is the gamma function) \cite{shevchenko10,gasparinetti11}.
The probability of the LZ transition is given by
$\LZ=e^{-2\pi\delta}$, where $\delta= E_J^2 \Time/ (8 E_C ) = \epsilon^2 E_C  \Time/ 2 $.
We consider the system initially in a superposition of ground and excited state and $|\psi_0\rangle= \alpha |g (0)\rangle + \sqrt{1-\alpha^2} \exp{(i \gamma)} |e (0)\rangle$.
We can then write for $t< \Time/2$, $ U(t) = U_1(t) $, while after the LZ transition at $t> \Time/2$, $ U(t) =U_2(t)  N_{LZ} U_1(\Time/2) $.
The calculation of $|\psi(t)\rangle $ is in the Supplementary Material (SM)\cite{SM}. Here we discuss only the relevant results.
\begin{figure}
    \begin{center}
   \includegraphics[scale=.8]{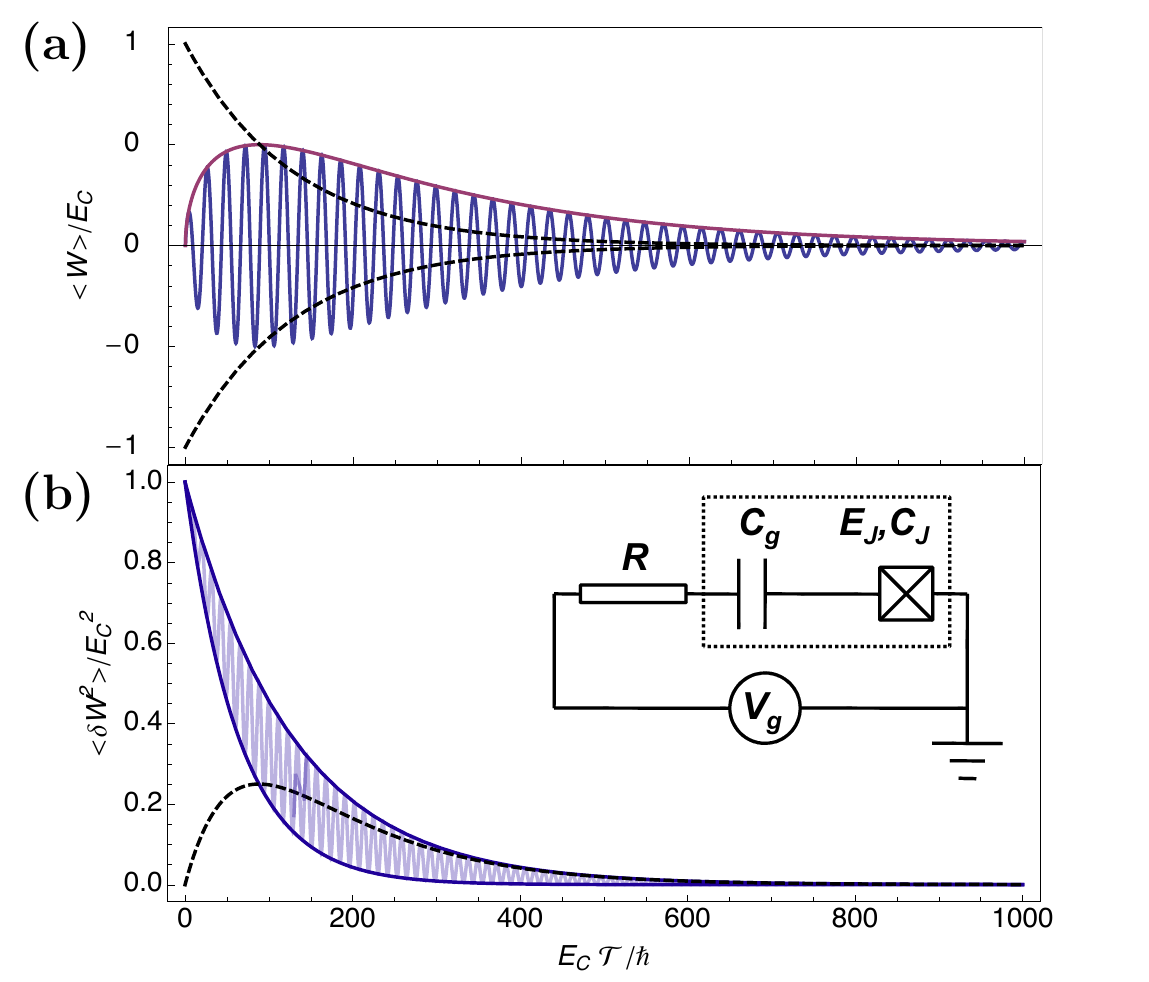}
   \end{center}
    \caption{(Color online) (a)
    Average work $\work$ normalized to $E_C$ for different initial states: $\alpha= 1/\sqrt{2}$ and $\gamma=0$ (blue oscillating curve), ground state $\alpha=1$ and $\gamma=0$ (top black exponentially decaying curve) and excited state $\alpha=0$ and $\gamma=0$ (lower black exponentially decaying curve). The purple curve denotes the behaviour $\sqrt{\LZ (1-\LZ)}$.
    (b) Work variance $\Wvariance$ normalized to $ E_C^2$. The blue solid lines contain the oscillatory behaviour for initial state $\alpha= 1/\sqrt{2}$ and $\gamma=0$.
    The black dashed line is obtained for initial state, $\alpha=1$ and $\gamma=0$ (ground state).
    We have used $\epsilon = E_J/(2E_C) =0.05$. 
    Inset: Circuit scheme of the Cooper-pair box (CPB) connected to a dissipative environment $R$.
    }
    \label{fig:Work_and_Variance}
\end{figure}

From Eq. (\ref{eq:work_CPB}), the corresponding average work is
\begin{eqnarray}
 && \work = E_C [(2 \alpha ^2-1) \LZ   \\ \nonumber
 && +2 \alpha  \sqrt{1-\alpha ^2} \sqrt{(1-\LZ) \LZ}
 \cos (\gamma +\varphi +2 \phid_1^{\frac{\Time}{2}}  ) ].
 \label{eq:AvW_nonad}
\end{eqnarray}
The first term represents the work done on the system which is initialized in the ground or in the excited state, i.e.,  $\alpha=1$ or $0$, respectively.
The second term with its characteristic oscillatory behavior is due to the quantum interference at the LZ avoided crossing \cite{gasparinetti11}.
This additional contribution is always present when the system is initially in a coherent state and it is a clear difference with the respect to the TMP. 
This difference is highlighted in Fig. \ref{fig:Work_and_Variance} $({\rm a})$ where we plot the analytical result in Eq. (\ref{eq:AvW_nonad}) as a function of $\Time$ for different initial states.
The oscillating behavior of $\work$ is obtained for $\alpha=1/\sqrt{2}$ and $\gamma=0$ and we should compare it with the prediction of the TMP $\work=0$.
The two exponential decays with $\LZ$ are obtained for ground and excited initial state, $\alpha=1$ and $\alpha=0$ with $\gamma=0$, respectively.
The behavior for thermalized initial density matrix can be obtained from this two curves with the correct weighted average.

With the same approach, the evaluation of Eq. (\ref{eq:work_variance}) yields $\Wsquare =   \LZ E_C^2$ for the second moment independent of the initial state.
With these results the corresponding rms fluctuation of work can be immediately calculated  as $\Wvariance = \Wsquare -\work ^2$.
Figure \ref{fig:Work_and_Variance} $({\rm b})$ shows the behavior of $\Wvariance$ for different initial states.
Numerical simulations confirm the analytical results presented in Figs. \ref{fig:Work_and_Variance}.

The definition of work in Eq. (\ref{eq:work_dynamics}) and the TMP give the same results if the system is initially in an eigenstate of $\hatH(0)$ or an incoherent superposition of them.
In the interesting case in which the system is initialized in the ground state, i.e., $\alpha=1$, and for nearly adiabatic drive ($\LZ\ll 1$),  we have a linear response result linking the average work and its fluctuations as $\Wvariance =   E_C \work$.

In this specific case, the first two moments of work calculated above agree with the full work distribution $\rho (W)$ which for a closed system with unitary evolution $U(\Time)$ can be found essentially by direct comparison of the initial, $\hat{H}(0)$, and final, $\hat{H}(\Time)$, Hamiltonian of the system. Indeed, the work generating function $G(u)$ (Fourier transform of the distribution) can be written as (see, e.g., \cite{talkner07}): 
$G(u) =  \mbox{Tr} \{ U^{\dagger} (\Time) e^{iu \hat{H}(\Time)} U(\Time) e^{-iu \hat{H}(0)} \rho_0 \}$
where $\rho_0$ is the initial density matrix of the system assumed to be diagonal together with the
initial Hamiltonian $\hat{H}(0)$. For the CPB considered above and the system initialized in the ground state, this equation gives $G(u) =  1+\LZ (e^{iu E_C}-1)$ which corresponds to the following work distribution:
\begin{equation}
\rho (W) = (1-\LZ) \delta(W) + \LZ \delta(W - E_C)\, .
\label{eq:dist2} \end{equation}
This distribution agrees with the first two moments 
and can be used to find the higher moments.

{\it Open system with slow and fast relaxation.}
The most interesting and non-trivial test of Eq. (\ref{eq:work_dynamics}) is when the system interacts with the environment during the evolution of $q$.
To evaluate the heat contribution in Eq. (\ref{eq:work_dynamics}) we need to consider a concrete example of the system-environment interaction.
If the  time and the ramp time are of the same order, dissipation $Q$ takes place during the driven evolution. To evaluate dissipation during the sweep, we then solve the master equation (ME) of the CPB adapted from the corresponding ME of Refs. \cite{pekola09,solinasPRB10}. This ME and some details of the analysis are given in the SM. The environment is described by the resistor $R$ coupled capacitively to the island of the CPB (inset in Fig. \ref{fig:Work_and_Variance}). As above, we assume that the temperature is low as compared to the excitation energy.
If the system is initially in the ground state, the average heat released to the environment during the ramp normalized by the total work done for a few values of $\epsilon$ is shown in Fig. \ref{fig:W_diss} (a) based on the numerical solution of the ME (solid lines). Dependence of the same quantity on the minimum energy gap $E_J$ (for scaling purposes the horizontal axis is $\epsilon^2$) is shown in (b). The apparent dependences on the various parameters in Fig. \ref{fig:W_diss} are captured by a simple analytical approximation
\begin{eqnarray} \label{approx}
Q/\langle W\rangle \simeq \frac{R}{R_Q}(\frac{C_g}{C_\Sigma})^2(\frac{E_J}{E_C})^2 \frac{E_C \Time}{\hbar},
\end{eqnarray}
which is derived in the SM with the assumption that, again, LZ-transition occurs exactly at $t=\Time/2$ with probability $P_{LZ}$. 
Here, $R_Q \equiv \hbar/e^2$. The prediction of Eq. \eqref{approx} is shown by dashed lines in Fig. \ref{fig:W_diss}, in close agreement with the full numerical solution. Energy relaxation occurs uniformly over the positive values of $q$ leading to proportionality of $Q/\langle W\rangle$ on $\Time$. As a by-product, Fig. \ref{fig:W_diss} (a) justifies the semi-quantitative analysis and the proposed measurement protocol above, since most of the work remains stored in the system during fast ramps (small $\Time$).
\begin{figure}
    \begin{center}
    \includegraphics[scale=.8]{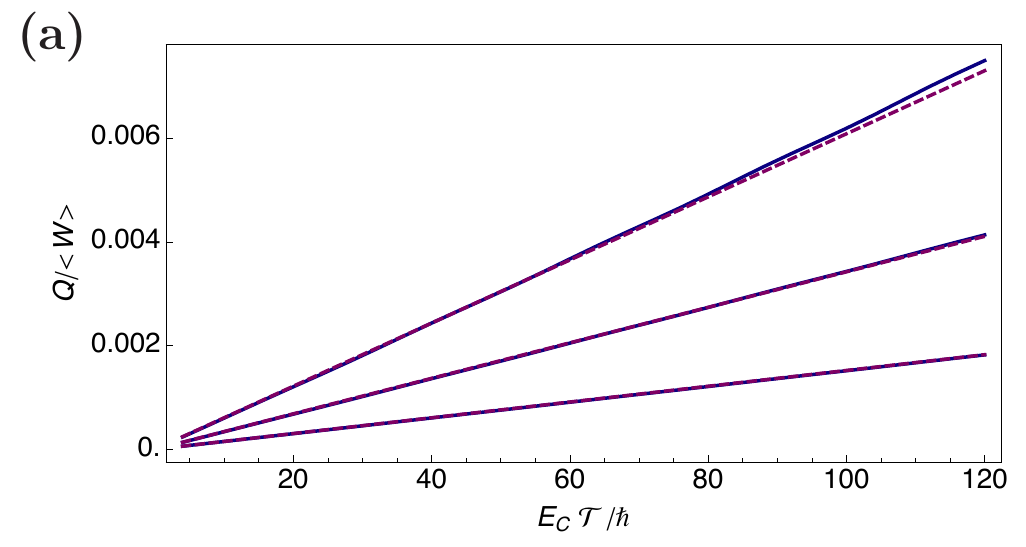}
     \includegraphics[scale=.8]{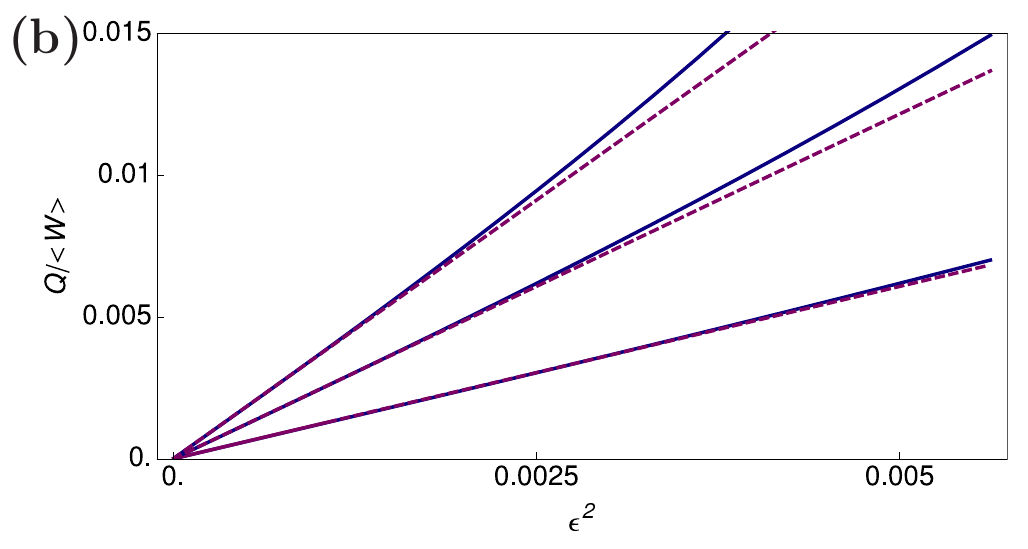}
   \end{center}
    \caption{(Color online) (a) Numerically calculated (solid lines) dissipated average heat during the sweep in an open CPB as a function of the sweep time when the system is initially in the ground state. The system-environment coupling constant is chosen to be $C_g/C_\Sigma=0.05$ here, $E_C/k_B = 1$ K, and the environment resistance is $R=1\cdot 10^4$ $\Omega$.
    The different curves from top to bottom correspond to $\epsilon =0.05$, $0.0375$, and $0.025$. The dashed line is the analytic approximation of Eq. \eqref{approx}. (b) Dissipated average heat at $E_C \Time/\hbar=150, 100$ and $50$ from top to bottom as a function of $\epsilon^2$. The other parameters and the line conventions are as in (a).}
    \label{fig:W_diss}
\end{figure}

In the limit of fast relaxation, i.e., $\tau \ll \Time$, it is possible to obtain an analytical estimate of the heat released during the evolution.
Since the dephasing time is usually smaller than the relaxation time, we can assume that the coherences between the ground and excited state are quickly lost and they do not influence the dynamics.
The system tends to follow the instantaneous equilibrium state but  due to the drive there are corrections to this dynamics which cause the heat emission. We calculate the heat in this regime directly from the semi-classical master equation. As shown in SM, the same result  
is still obtained if one takes into account quantum corrections from the finite drive to adiabatic dynamics. 

The integrand in the heat contribution in Eq. (\ref{eq:work_dynamics}) can be conveniently written as $ -\hbar \dot{\rho}_{gg} \omega_0$.
If we denote with $\Gamma_{ge}$ and $\Gamma_{eg}$ the excitation and relaxation rates, respectively,
in the semi-classical limit we have $\dot{\rho}_{gg} = -\Gamma_\Sigma \rho_{gg} + \Gamma_{ge}$ where $\Gamma_\Sigma = \Gamma_{ge}+\Gamma_{eg}$, $\Gamma_{ge}/\Gamma_{eg} = \exp{(-\beta \hbar \omega_0)}$ and $\beta$ is the inverse temperature of the environment.
We use the trial solutions $\rho_{gg} = \rho_{gg}^{(0)} + \delta  \rho_{gg}^{(1)}$ where $ \rho_{gg}^{(0)} = \Gamma_{ge}/\Gamma_{\Sigma}$ is the stationary solution and $\delta  \rho_{gg}^{(1)}$ is the correction due to the drive.
Plugging $\rho_{gg}$ in the initial equation we obtain $ \delta  \rho_{gg}^{(1)} = -\dot{\rho}_{gg}^{(0)}/\Gamma_\Sigma $.
When integrated the adiabatic contribution $\hbar \dot{\rho}_{gg}^{(0)}\omega_0$ gives no contribution (see SM) and, with an integration by parts, the non-vanishing contribution in the limit $\beta E_C \gg 1$ can be written as
\begin{eqnarray}
 Q &=& \frac{4 E_C^2 }{\hbar \Time} \int_{-\frac{1}{2}}^{\frac{1}{2}} dq  \frac{q}{\omega_0 \Gamma_\Sigma} \frac{d}{dq} \Big( \frac{\Gamma_{eg}}{\Gamma_\Sigma}\Big) \nonumber \\
 &=& \frac{\beta E_C^2}{\Time} \int_{-\frac{1}{2}}^{\frac{1}{2}} dq  \frac{\eta^2}{\Gamma_\Sigma \cosh^2(\frac{\beta \omega_0}{2})}
\end{eqnarray}
Here the second line arises from the detailed balance of the transition rates.
Thus, we recover the expected properties of the released heat: $(i)$ it depends on the full evolution which, in this limit, is represented by the driving parameter $q$, $(ii)$ it  scales as $1/\Time$ and $(iii)$ it is positive \cite{sekimoto}.

In summary, we have analyzed work done by a driving field on a quantum system. The obtained expression of average work has a physical interpretation allowing one to assign separate contributions to the change in the internal energy and the heat dissipated to the environment in the spirit of the first law of thermodynamics.
We applied our results to a two-level system obeying in the first case unitary evolution and then in the presence of weak dissipation. For an open system, we presented a detailed analysis of the released heat two regimes where the relaxation time was either of the order or smaller than the driving time.
In the latter case, our approach allows an analytical calculation of the released heat which has and immediate physical interpretation.



We would like to thank T. Ala-Nissil\"a, A. Kutvonen, S. Suomela, S. Gasparinetti, M. M\"ott\"onen and J. Ankerhold for useful discussions.
This work was supported by the European Community FP7 under grants No. 238345 GEOMDISS and Academy of Finland Centre of Excellence.
P.S. acknowledges financial support from FIRB Ð Futuro in Ricerca 2012 under Grant No. RBFR1236VV ÒHybridNanoDevÓ.


\begin{widetext}

\date{\today}

\maketitle

\appendix{}

\section{Dynamical evolution in the instantaneous Landau-Zener transition model}

To make the Landau-Zener problem analytically manageable, we assume that the transition occurs instantaneously at time $t=\Time/2$ \cite{app_shevchenko10, app_gasparinetti11}.
In this case, the unitary evolution operator can be written as $ U(t) = U_1(t) $ for $t< \Time/2$, and as $U(t) =U_2(t)  N_{LZ} U_1(\Time/2) $ after the LZ transition at $t> \Time/2$.
If the system is initially in the state $|\psi_0\rangle= \alpha |g(0)\rangle + \beta \exp{(i \gamma)} |e(0)\rangle$, we obtain
\begin{eqnarray} \label{eq:LZ_ev1}
 |\psi(t)\rangle =
         e^{-i \phid_1^t} \alpha |g(t)\rangle +\beta e^{i (\gamma+ \phid_1^t)} |e(t)\rangle
         \end{eqnarray}
 for $t < \Time/2$, and
 \begin{eqnarray} \label{eq:LZ_ev2}
 |\psi(t)\rangle &=&   (\alpha  \sqrt{1-\LZ}  e^{-i (\phid_1^{\frac{\Time}{2}}+i \phid_2^t+i
   \varphi )}-\beta  \sqrt{\LZ} e^{i (\gamma + \phid_1^{\frac{\Time}{2}} - \phid_2^t)}  )|g(t)\rangle \nonumber \\
   &+&
   (\alpha  \sqrt{\LZ} e^{i \phid_2^t-i \phid_1^\frac{\Time}{2}}+\beta  \sqrt{1-\LZ} e^{i \gamma +i \phid_1^\frac{\Time}{2}+i \phid_2^t+i \varphi }) |e(t)\rangle
\end{eqnarray}
for $\Time/2 < t < \Time$.

If the Cooper pair sluice is in the charging regime, i.e., $E_C/E_J \ll 1$, the $\eta $ function is well approximated by the {\it signum} function up to a correction of order $\epsilon^2$: $\eta(t) =-1$ if $t < \Time/2$ and $\eta(t) =+1$ if $t > \Time/2$.
This implies that the ground and the excited states can be approximated by the charge states as
$\{|g(t)\rangle, |e(t)\rangle\} = \{ |0\rangle,  |1\rangle \}$ for $t < \Time/2$ and $\{|g(t)\rangle, |e(t)\rangle\} = \{ |1\rangle,  |0\rangle \}$ for $t > \Time/2$. This allows us to calculate directly the quantities of interest as
\begin{equation}
  \langle \hatn(t) \rangle=
  \left \{ \begin{array}{cc}
 1- \alpha^2 &~{\rm for~ } 0 < t < \frac{\Time}{2} \\
 (1- 2 \alpha^2) \LZ +\alpha (\alpha -2 \sqrt{1- \alpha^2} \cos(\gamma +\varphi+ \phid_1^\frac{\Time}{2}) \sqrt{(1-\LZ)\LZ})  &~{\rm for~ } \frac{\Time}{2} < t < \Time \\
\end{array}
\right .\
\end{equation}
Since  $\langle \hatn(t) \rangle$ does not depends on time the integration is trivial and we obtain immediately the results of Eq. $(8)$ in the main text.

For the two-point correlator, it is convenient use the equivalent definition
$\int_0^\Time dt_2 \int_0^{\Time} dt_1   \langle \hatn^H(t_2) \hatn^H(t_1)\rangle  = 2 \int_0^\Time dt_2 \int_0^{t_2} dt_1    \Re{\rm e}( \langle \hatn^H(t_2) \hatn^H(t_1)\rangle)$ which is obtained simply using the symmetric property of the integral and the fact that $(\langle \hatn^H(t_2) \hatn^H(t_1)\rangle)^\dagger = \langle \hatn^H(t_2) \hatn^H(t_1)\rangle$.
Then, we write
\begin{equation}
\Re{\rm e}( \langle\hatn^H(t_2) \hatn^H(t_1)\rangle)  = \Re{\rm e}(\langle\psi_0|  \hatU^\dagger(t_2) |1\rangle  \langle 1| \hatU(t_2) \hatU^\dagger(t_1) |1\rangle \langle 1| \hatU(t_1)   |\psi_0 \rangle )
\end{equation}
which can be calculated in a similar way.

It turns out that, $\Re{\rm e}( \langle\hatn^H(t_2) \hatn^H(t_1)\rangle)$ does not depend on time and
\begin{equation}
  \int_0^\Time dt_2 \int_0^{t_2} dt_1 \Re{\rm e} [ \langle \hatn^H(t_2) \hatn^H(t_1)\rangle] =
  \left \{ \begin{array}{ll}
 \frac{\Time^2}{8} \left(1-\alpha ^2\right)  \\
 \frac{\Time^2}{4} \left[\left(1-\alpha ^2\right) \LZ-\alpha  \sqrt{\left(\alpha
   ^2-1\right) \left(\LZ-1\right) \LZ} \cos \left(\gamma +2 \phid_1^\frac{\Time}{2}+\varphi \right)\right]  \\
   \frac{\Time^2}{8} \left[ \left(1-2 \alpha ^2\right)
   \LZ + \alpha  \left(\alpha -2 \sqrt{\left(1-\alpha ^2\right)
   \left(1-\LZ \right) \LZ} \cos \left(\gamma +2 \phid_1^\frac{\Time}{2}+\varphi \right)\right)\right]
\end{array}
\right .\
\end{equation}
where the three contributions come from the integration intervals $0 < t_1 < t_2 < \frac{\Time}{2}$, $0 < t_1 < \frac{\Time}{2} < t_2 < \Time$ and $0  < \frac{\Time}{2}< t_1 < t_2 < \Time$, respectively.
From Eqs. $(6)$ and $(8)$ in the main text, 
the second moment is obtained as
\begin{equation}
\Wsquare =E_C^2 \LZ.
\end{equation}

\section{Behavior of the heat in the weak coupling regime}

The full master equation that we solve numerically reads \cite{me10}
\begin{eqnarray} \label{fm1}
&&\dot \rho_{gg} = -2v_{ge}\Re{\rm e} (\rho_{ge})-(\Gamma_{ge}+\Gamma_{eg})\rho_{gg}+\Gamma_{eg}+\tilde\Gamma_0 \Re{\rm e} (\rho_{ge})\nonumber \\&&
\dot\rho_{ge} = v_{ge}(2\rho_{gg}-1)+i\omega_0\rho_{ge}-i(\Gamma_{ge}+\Gamma_{eg})\Im{\rm m}(\rho_{ge})-\Gamma_\varphi\rho_{ge}+(\tilde\Gamma_++\tilde \Gamma_-)\rho_{gg}-\tilde\Gamma_+\nonumber \\ && +i\frac{v_{ge}}{\omega_0}\big[(\Gamma_{eg}-\Gamma_{ge})-2(\Gamma_++\Gamma_-)\rho_{gg}+2\Gamma_++\Gamma_\varphi(2\rho_{gg}-1)+2(\tilde \Gamma_0-\tilde \Gamma_+-\tilde\Gamma_-)\Re{\rm e}(\rho_{ge})\big].
\end{eqnarray}
Here
\begin{equation} \label{vge}
v_{ge}=\frac{1}{2}\frac{\epsilon}{q^2+\epsilon^2}\,\dot {q}
\end{equation}
is the driving term, and the various rates related to the interaction with the environment read
$\Gamma_{ge}=\frac{m_2^2}{\hbar^2}S_{V_g}(-\omega_{0})$,
$\Gamma_{eg}=\frac{m_2^2}{\hbar^2}S_{V_g}(+\omega_{0})$,
$\Gamma_{\varphi}=2\frac{m_1^2}{\hbar^2}S_{V_g}(0)$,
$\tilde{\Gamma}_\pm=\frac{m_1m_2}{\hbar^2}S_{V_g}(\pm\omega_{0})$,
$\tilde{\Gamma}_0=2\frac{m_1m_2}{\hbar^2}S_{V_g}(0)$, $\Gamma_\pm=\frac{m_1^2}{\hbar^2}S_{V_g}(\pm\omega_{0})$, and $\Gamma_0=2\frac{m_2^2}{\hbar^2}S_{V_g}(0)$. The couplings are defined as $m_1= -\eta eC_g/C_\Sigma$ and $m_2=\sqrt{1-\eta^2}eC_g/C_\Sigma$, and $S_{V_g}(\omega)$ is the noise spectrum.
In the numerical solution we have assumed temperature to be zero.
With $\eta=q/\sqrt{q^2+\epsilon^2}$, $S(+\omega_0)=2R\hbar\omega_0$, and
\begin{equation} \label{qheat4a}
\hbar \omega_0=2E_C\sqrt{q^2+\epsilon^2},
\end{equation}
we obtain the relaxation rate as
\begin{equation} \label{qheat4}
\Gamma_{eg}=4\frac{R}{R_Q}(\frac{C_g}{C_\Sigma})^2\frac{\epsilon^2}{\sqrt{q^2+\epsilon^2}}\frac{E_C}{\hbar},
\end{equation}
where $R_Q =\hbar/e^2$.

We consider the case in which the system is initially in the ground state and $\rho_{gg}(0) = 1$.
For the analytic approximation, we assume that at $t=\Time/2$ when the system passes the degeneracy, the population of the excited state $\rho_{ee}=1-\rho_{gg}$ jumps from $0$ to $P_{LZ}$. (The ramp starts at $t=0$ and ends at $t=\Time$.) After $t=\Time/2$, the excited state population relaxes approximately as
\begin{eqnarray} \label{qheat5}
\dot \rho_{ee} = -\Gamma_{eg}\rho_{ee},
\end{eqnarray}
which yields
\begin{eqnarray} \label{qheat6}
\rho_{ee}(t) =\rho_{ee}(0)e^{-\int_{\Time/2}^t \Gamma_{eg}(\tau)d\tau}\simeq \rho_{ee}(0)[1-\int_{\Time/2}^t \Gamma_{eg}(\tau)d\tau],
\end{eqnarray}
where in the last step we have assumed that the relaxation is weak on the time scale of the sweep, $\Gamma_{eg}\Time \ll 1$. The dissipated heat in the ramp can be approximated by $Q\simeq \int_{\Time/2}^\Time \dot \rho_{gg}(\tau) \hbar\omega_0(\tau) d\tau$. Then we have by inserting \eqref{qheat4a}, \eqref{qheat4} and \eqref{qheat6} into the expression of $Q$:
\begin{eqnarray} \label{qheat7}
Q\simeq 4P_{LZ} E_C \frac{R}{R_Q}(\frac{C_g}{C_\Sigma})^2\epsilon^2 \frac{E_C \Time}{\hbar},
\end{eqnarray}
and by dividing by the average work $\langle W\rangle \simeq P_{LZ}E_C$, we have
\begin{eqnarray} \label{qheat8}
Q/\langle W\rangle \simeq \frac{R}{R_Q}(\frac{C_g}{C_\Sigma})^2(\frac{E_J}{E_C})^2 \frac{E_C \Time}{\hbar},
\end{eqnarray}
which is Eq. $(10)$ of the main text.

\section{Semiclassical approximation of the heat in the fast relaxation regime}

In the fast relaxation regime, $\tau \ll \Time$, when the system is driven adiabatically we must include in the master equation the effect of the drive. It can be consider as a perturbation to the adiabatic dynamics \cite{me10}.
This perturbation is given by the terms $v_{ge}/\omega_0$ [where $v_{ge}$ is given in Eq. (\ref{vge}) ] which we assume to be small  during the evolution.
As a rough estimate of the adiabatic parameter we can use $\chi=\hbar /(2 \Time E_J)$ since $ 2 E_J$ is the smallest energy gap at the Landau-Zener crossing.

At the zero-th order in $\chi$ the master equation reads
\begin{equation}
 \dot{\rho}_{gg}^{(0)} = - \Gamma_\Sigma^{(0)} \rho_{gg}^{(0)} +\Gamma_{eg}^{(0)}
\end{equation}
with $\rho_{ge}^{(0)} =0$.
The quasi-stationary (adiabatic) solution is $\rho_{gg}^{(0)}  = \Gamma_{eg}^{(0)} / \Gamma_\Sigma^{(0)} = 1/(1+e^{-\beta \hbar \omega_0})$ if we assume that the detailed balance $\Gamma_{ge}/\Gamma_{eg} = \exp{(-\beta \hbar \omega_0)}$ is instantaneously satisfied.
The heat contribution reads explicitly
\begin{equation}
 Q^{(0)}= \int_0^\Time dt \hbar \omega_0 \dot{\rho}_{gg}^{(0)}=\frac{\hbar \beta }{4}\int_{-E_C/2}^{E_C/2} d\omega_0  \frac{\omega_0}{ \cosh^2 ( \frac{\beta \omega_0}{2})}=0.
\end{equation}

Then we need to calculate the correction to the first order in $\chi$.
This can be done starting from the master equation written in the superadiabatic basis $\{ |g^{(1)}\rangle, |e^{(1)}\rangle \}$\cite{salmilehto11} which allows us to include directly the high order correction in $\chi$ in this basis.
Under the hypothesis of fast dephasing we have
\begin{equation}
 \dot{\rho}_{gg}^{(1)} = - \Gamma_\Sigma^{(1)} \rho_{gg}^{(1)} +\Gamma_{eg}^{(1)}.
 \label{eq:rho_1}
\end{equation}
Formally we can expand the functions in the above equation as
\begin{eqnarray}
 \rho_{gg}^{(1)} = \rho_{gg}^{(0)} + \delta \rho_{gg}^{(1)} \nonumber \\
  \Gamma_\Sigma^{(1)} = \Gamma_\Sigma^{(0)} + \delta \Gamma_\Sigma^{(1)} \nonumber \\
    \Gamma_{eg}^{(1)} = \Gamma_{eg}^{(0)} + \delta \Gamma_{eg}^{(1)}.
\end{eqnarray}
Using the explicit expression for the $\rho_{gg}^{(0)}$ in Eq. (\ref{eq:rho_1}) and taking the terms of the same order in $\chi$ we obtain
\begin{equation}
\dot{\rho}_{gg}^{(0)} = - \Gamma_\Sigma^{(0)}  \delta \rho_{gg}^{(1)} - \frac{ \delta \Gamma_\Sigma^{(1)}  \Gamma_{eg}^{(0)} }{\Gamma_\Sigma^{(0)} }+\delta \Gamma_{eg}^{(1)}
\end{equation}
which gives
\begin{equation}
\delta \rho_{gg}^{(1)}  = -\frac{\dot{\rho}_{gg}^{(0)}}{\Gamma_\Sigma^{(0)}} -\frac{ \delta \Gamma_\Sigma^{(1)}  \Gamma_{eg}^{(0)}}{[\Gamma_\Sigma^{(0)}]^2  }+\frac{ \delta \Gamma_{eg}^{(1)} }{\Gamma_\Sigma^{(0)} }
\end{equation}
This equation is at the first order in the adiabatic parameter $\chi$ and the remaining task is to calculate the perturbation in the rates $\delta \Gamma_i^{(1)}$.
We have \cite{salmilehto11}
\begin{equation}
 \Gamma_{eg}^{(1)} = \frac{|\langle e^{(1)} | \hat{A} |g^{(1)} \rangle|^2}{\hbar^2} S(\omega_0^{(1)})
 \label{eq:Gammas}
\end{equation}
where $ \hat{A}$ is the noise operator acting on the system, $S(\omega_0^{(1)})$ is the spectral density function and $\omega_0^{(1)}$ is the energy renormalized for the drive contribution.
By direct calculation and at the first order in the adiabatic parameter $\chi$, we have that $\omega_0^{(1)} = \omega_0$, $|g^{(1)} \rangle = |g \rangle - i v_{ge}/\omega_0 |e \rangle$ and $|g^{(1)} \rangle = |e \rangle + i v_{ge}/\omega_0 |g \rangle$.
For the Cooper pair box the dominant noise is the charge noise which is described by  $\hat{A} = | 1\rangle \langle 1| - | 0\rangle \langle 0|$.
Expanding Eq. (\ref{eq:Gammas}) in $\chi$, we can check directly that there is no correction to the transition rates, i.e., $\delta \Gamma_i^{(1)}=0$, and,  finally we obtain
\begin{equation}
 \delta \rho_{gg}^{(1)}  = -\frac{\dot{\rho}_{gg}^{(0)}}{\Gamma_\Sigma^{(0)}}
\end{equation}
as in the main text.
With this result the released heat explicitly reads $Q=- \int_0^{\Time} \hbar \omega_0 \frac{d}{dt } \Big[ \frac{1}{\Gamma_\Sigma} \frac{d}{dt} \Big( \frac{\Gamma_{eg}}{\Gamma_\Sigma}\Big) \Big]$.
Integrating by parts, exploiting the detailed balance condition and the symmetry $\omega_0(\Time) =\omega_0(0)= E_C$, the first contribution reads
\begin{equation}
  -\frac{\beta}{2}  \frac{\hbar^2 \omega_0(t) \dot{\omega}_0(t)}{\Gamma_{\Sigma} \cosh^2 (\frac{\beta \omega_0(t) }{2})} \Big|_0^{\Time}.
\end{equation}
which vanishes exponentially in the limit $\beta E_C \gg 1$.
Thus, the only remaining contribution is the one in Eq. $(11)$ of the main text.

\end{widetext}

\end{document}